\documentclass[3p,times,procedia]{elsarticle}
\usepackage{nupha_ecrc}

\volume{00}

\firstpage{1}

\journalname{Nuclear Physics A}

\runauth{}

\jid{nupha}

\jnltitlelogo{Nuclear Physics A}

\usepackage{amssymb}

\biboptions{square,comma,numbers,sort&compress}

\usepackage[figuresright]{rotating}

\usepackage{graphicx}
\usepackage{dcolumn}
\usepackage{bm}
\usepackage{amssymb}
\usepackage{xspace}

\usepackage{hyperref}
\usepackage{amsmath}

\usepackage{color}

\newcommand{\pp}{\ensuremath{\rm pp}\xspace}

\newcommand{\pt}{\ensuremath{p_{\rm{T}}}\xspace}

\newcommand{\dedx}{\ensuremath{{\rm d}E/{\rm d}x}\xspace}

\newcommand{\pbpb}{Pb--Pb\xspace}
\newcommand{\auau}{Au--Au\xspace}
\newcommand{\ppb}{p-Pb\xspace}
\newcommand{\Rppb}{\ensuremath{R_{\rm pPb}}\xspace}

\newcommand{\mpt}{\ensuremath{\langle p_{\rm T} \rangle \xspace}}

\newcommand{\snnt}[1]{\ensuremath{\sqrt{s_{\rm NN}} = #1 \text{\,TeV}}\xspace}

\newcommand{\sppt}[1]{\ensuremath{\sqrt{s} = #1 \text{\,TeV}}\xspace}

\begin{document}

\begin{frontmatter}



\dochead{}

\title{Light flavor results in p-Pb collisions with ALICE}


\author{Antonio Ortiz \\ on behalf of the ALICE Collaboration}

\address{Instituto de Ciencias Nucleares, Universidad Nacional Aut\'onoma de M\'exico. \\ Circuito exterior s/n, Ciudad Universitaria, Del. Coyoac\'an, C.P. 04510, M\'exico D. F.}

\begin{abstract}

Particle ratios provide insight into the hadrochemistry of the event and the mechanisms for particle production. In Pb-Pb collisions the relative multi-strange baryon yields exhibit an enhancement with respect to pp collisions, whereas the short-lived K$^{*0}$ resonance is suppressed in the most central events due to re-scattering of its decay daughter particles. Measurements in p-Pb allow us to investigate the development of these effects as a function of the system size.

We report comprehensive results on light-flavor hadron production measured with the ALICE detector in p-Pb collisions at $\sqrt{s_{\rm NN}}=5.02$\,TeV, covering a wide range of particle species which includes long-lived hadrons, resonances and multi-strange baryons. The measurements include the transverse momentum spectra and the ratios of spectra among different species, and extend over a very large transverse momentum region, from $\approx$100\,MeV/$c$ to $\approx$20\,GeV/$c$, depending on the particle species.

\end{abstract}


\end{frontmatter}


\section{Introduction}
\label{sec:1}

In central heavy-ion collisions at ultra relativistic energies it is well established that a strongly interacting medium of quarks and gluons is formed (sQGP).  The system observed in $\sqrt{s_{\rm NN}}=0.2$\,TeV \auau collisions at the Relativistic Heavy Ion Collider (RHIC) was found to behave as a perfect fluid which is initially close to the ideal hydrodynamic limit~\cite{Adams:2005dq,Adcox:2004mh,Back:2004je,Arsene:2004fa}. The main features of this new form of matter are the strong flow (radial and anisotropic) and the opacity to jets. The strongly coupled QGP has been confirmed and studied by experiments at the Large Hadron Collider (LHC)~\cite{Schukraft:2011na}, where \pbpb collisions at $\sqrt{s_{\rm NN}}=2.76$\,TeV have been achieved~\cite{Heinz:2011kt}.

To disentangle the so-called cold nuclear matter effects from those attributed to sQGP, data from control experiments like \pp and \ppb collisions are also analyzed. Surprisingly, results from the LHC showed that high multiplicity \pp and \ppb collisions exhibit characteristics reminiscent of those due to final state effects (flow-like patterns and the ridge structure), but no sign of jet quenching~\cite{ABELEV:2013wsa,Abelev:2013haa}. The main focus of this work is to present a review of recent measurements on light flavor identified particle production ($\pi^{+}$+\,$\pi^{-}$, K$^{+}$+\,K$^{-}$, K$^{0}_{\rm S}$, ${\rm p}$+$\bar{\rm p}$,  $\phi$, K$^{*0}$,  ${\Lambda}$+\,$\bar{\Lambda}$, $\Xi^{-}$+\,$\bar{\Xi}^{+}$ and $\Omega^{-}$+\,$\bar{\Omega}^{+}$) in \ppb collisions at $\sqrt{s_{\rm NN}}=5.02$\,TeV to contribute to the understanding of the sQGP-like effects in small systems~\cite{Schenke:2015aqa,Bozek:2014era,Ortiz:2013yxa,Armesto:2015ioy,Bautista:2015kwa,Bierlich:2014xba}.

\section{Data analysis}
\label{sec:2}

The results discussed in the present paper are obtained from the analysis of the \ppb data recorded with the ALICE (A Large Ion Collider Experiment) detector~\cite{Aamodt:2008zz,Abelev:2014ffa} at the LHC during 2013. The minimum bias  (MB) trigger signal was provided by the V0 counters, two arrays of 32 scintillator tiles each covering the full azimuth within 2.8$<$\,$\eta_{\rm lab}$\,$<$5.1 (V0A, Pb beam direction) and -3.7$<$\,$\eta_{\rm lab}$\,$<$-1.7 (V0C, p beam direction). The signal amplitude and arrival time collected in each tile were recorded. A coincidence of signals in both V0A and V0C detectors was required to remove contributions from single diffractive and electromagnetic events.
For \ppb collisions, the nucleon-nucleon center-of-mass system has a rapidity of $y_{\rm NN}$=-0.465 in the direction of the proton beam. Therefore, to ensure good detector acceptance, identified particle measurements shown in this paper are performed within -0.5$<$\,$y_{\rm CMS}$\,$<$0.0 in the nucleon-nucleon center-of-mass system, unless specified otherwise. Particle production is studied in seven event classes being defined based on the total charge deposited in the V0A detector~\cite{Abelev:2013haa}. It is worth noting that in \ppb the correlation of geometry and multiplicity is much weaker than in \pbpb and hence the event classes are not denoted as centrality classes in this work.


Particle IDentification (PID) of charged pions, kaons and (anti)protons in ALICE is done exploiting all known techniques. Hadrons at low momenta (from $\approx$100-300\,MeV/$c$ up to $\approx$2-3\,GeV/$c$) can be identified track-by-track using different detectors. The specific energy loss (\dedx) in gas and in silicon is measured in the Time Projection Chamber (TPC) and the Inner Tracking System (ITS), respectively.  The TOF detector allows for particle identification measuring the particle velocity with the Time-Of-Flight technique. At higher \pt ($>$2-3\,GeV/$c$ up to 20\,GeV/$c$), PID is done measuring the TPC-\dedx in the relativistic rise regime of the Bethe-Bloch curve, where the TPC-\dedx separation between particles with different masses is constant at large momenta~\cite{Adam:2015kca}. At intermediate \pt (2$<$\,$\pt$\,$<$6 GeV/$c$) the PID is improved using a High Momentum Cherenkov PID detector (HMPID). K$^{*0}$ and $\phi$ signals are reconstructed following the techniques which are described in~\cite{Abelev:2012hy,Abelev:2014uua}. For each event, the invariant-mass distribution of the K$^{*0}$ and $\phi$ is constructed using all unlike-sign combinations of charged kaon candidates with charged pion and kaon candidates, respectively. Daughter tracks are identified using the TPC and TOF detectors. The K$^{0}_{\rm S}$ and $\Lambda$ particles are identified by reconstructing the weak decay topology in the channels ${\rm K}^{0}_{\rm S}$\,$\rightarrow$\,$\pi^{+} \pi^{-}$ and $\Lambda$($\bar{\Lambda}$)\,$\rightarrow$\,${\rm p} \pi^{-}$(${\bar{\rm p}}\pi^{+}$). Similarly for multi-strange baryons, topological PID is performed reconstructing the decay channels $\Xi^{-}$\,$\rightarrow$\,$\Lambda \pi^{-}$ and $\Omega^{-}$\,$\rightarrow$\,$\Lambda$K$^{-}$ and the subsequent $\Lambda$ weak decay.

\begin{figure}[htbp]
\begin{center}
   \includegraphics[width=0.45\textwidth]{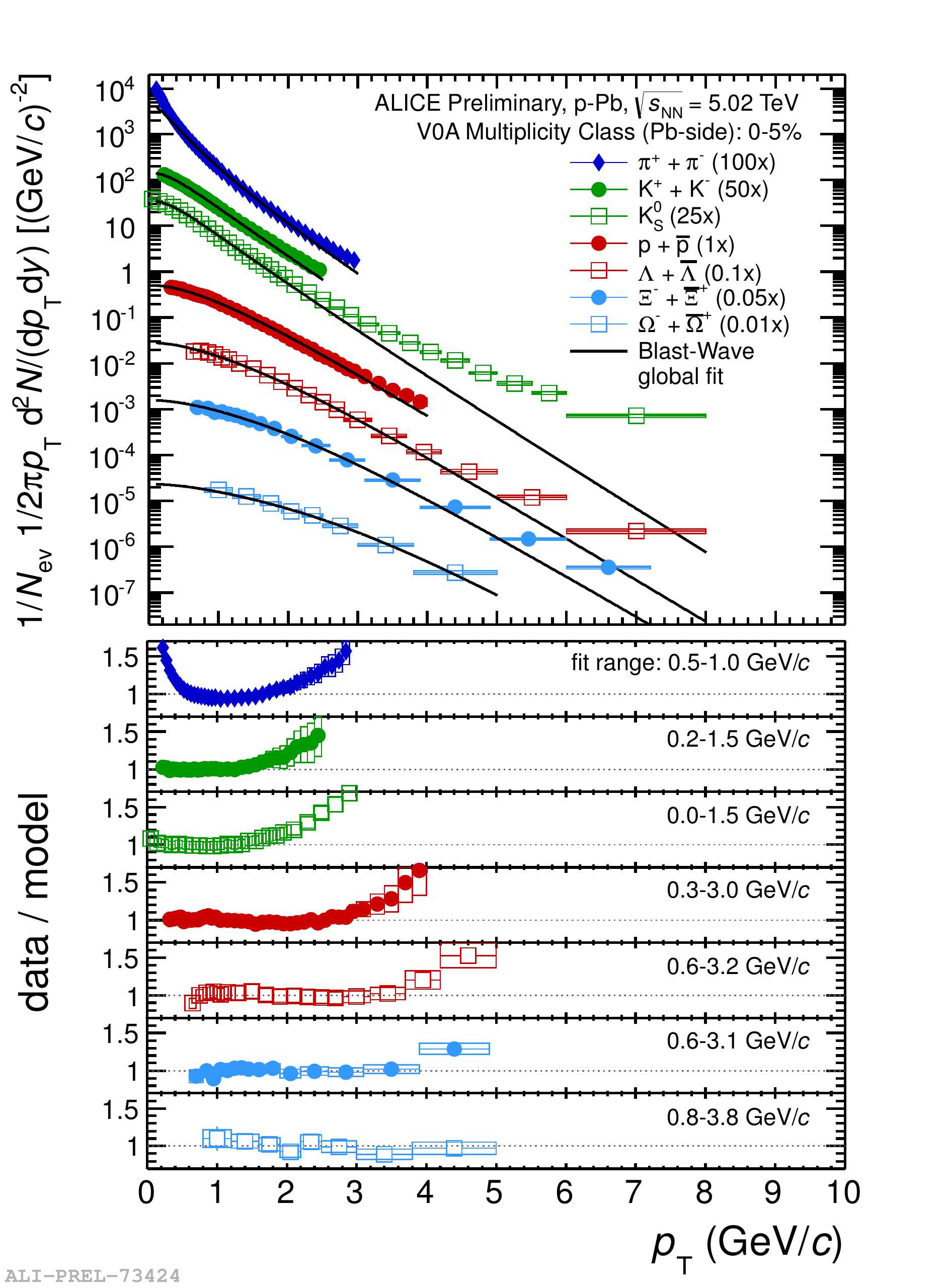}
   \includegraphics[width=0.53\textwidth]{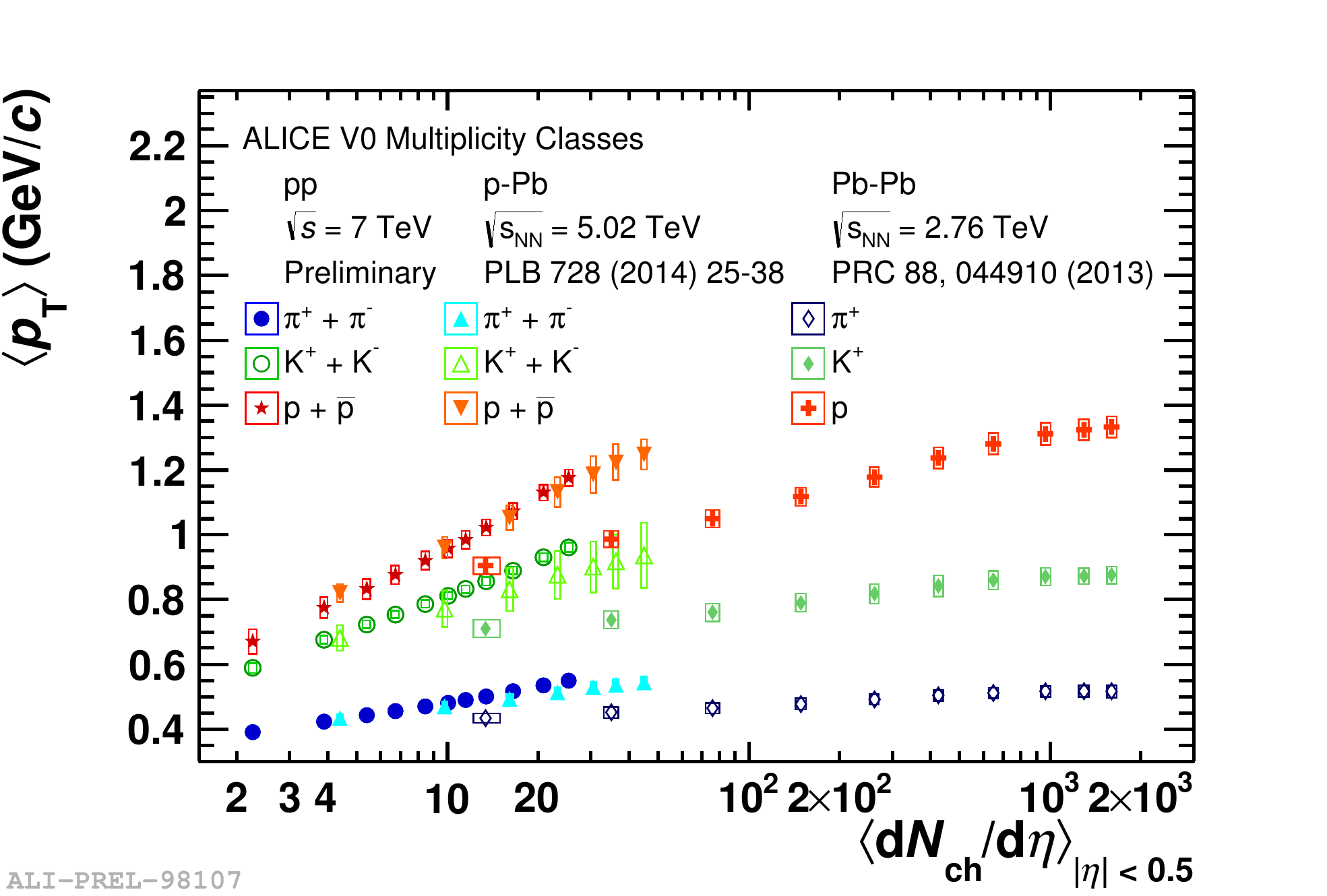}   
   \caption{(Color online).  Transverse momentum distributions of ($\Xi$) and ($\Omega$) particles in the 0-5\% V0A multiplicity class  are compared to the blast wave prediction obtained from the simultaneous blast-wave fit to the pion, kaon, proton and lambda \pt spectra (left). Average \pt as a function of the average multiplicity at mid-rapidity for different colliding systems. Results for charged pions, kaons and (anti)protons are shown (right).}
  \label{fig:pPb:1}
\end{center}
\end{figure}

\section{Results}
\label{sec:2}

The transverse momentum spectra of charged pions, kaons and (anti)protons as a function of the event multiplicity have been measured up to 15\,GeV/$c$. At low \pt ($<$\,$2$-$3$\,GeV/$c$) the spectra exhibit a hardening with increasing multiplicity, with this effect being more pronounced for heavy particles. We are therefore observing features which resemble the radial flow effects well know from heavy-ion collisions~\cite{Adam:2015kca} and which are well described when a hydrodynamical evolution of the system is considered.  In~\cite{Abelev:2013haa}  it was shown that for high multiplicity \ppb events, the \pt spectra were described by the blast-wave function which is a simplified hydrodynamical evolution model. Using the parameters obtained from the simultaneous fit to pion, kaon, proton and lambda \pt spectra the model is able to describe the multi-strange baryon \pt distributions ($\pt$\,$<$4\,GeV/$c$) as shown in Fig.~\ref{fig:pPb:1}. The feature is also observed in Pythia 8 tune 4C~\cite{Corke:2010yf,Ortiz:2013yxa}, where no hydrodynamical evolution is modeled. To compare the different colliding systems (\pp, \ppb and \pbpb collisions), the right hand side of Fig.~\ref{fig:pPb:1} shows the average \pt as a function of $\langle {\rm d}N_{\rm ch}/ \eta \rangle_{|\eta|<0.5}$ for charged pions, kaons and (anti)protons.  The V0A detector is also used to classify the \pp collisions according with their multiplicity. For small systems, the average \pt exhibits a steeper rise with increasing multiplicity than that for \pbpb collisions. It is also observed that \mpt\,increases faster with multiplicity for heavier hadrons than for lighter ones. Additionally, it has been pointed out that the average \pt shows an explicit dependence with the number of constituent quarks in: 1) MB \pp collisions at different $\sqrt{s}$, 2) the different multiplicity classes reported for \ppb collisions and 3) for peripheral \pbpb collisions (60-90\%). This feature has been attributed to fragmentation~\cite{Ortiz:2015cma} and is not present in the hydrodynamical calculations~\cite{Bozek:2013ska}. 

\begin{figure}[htbp]
\begin{center}
   \includegraphics[width=0.52\textwidth]{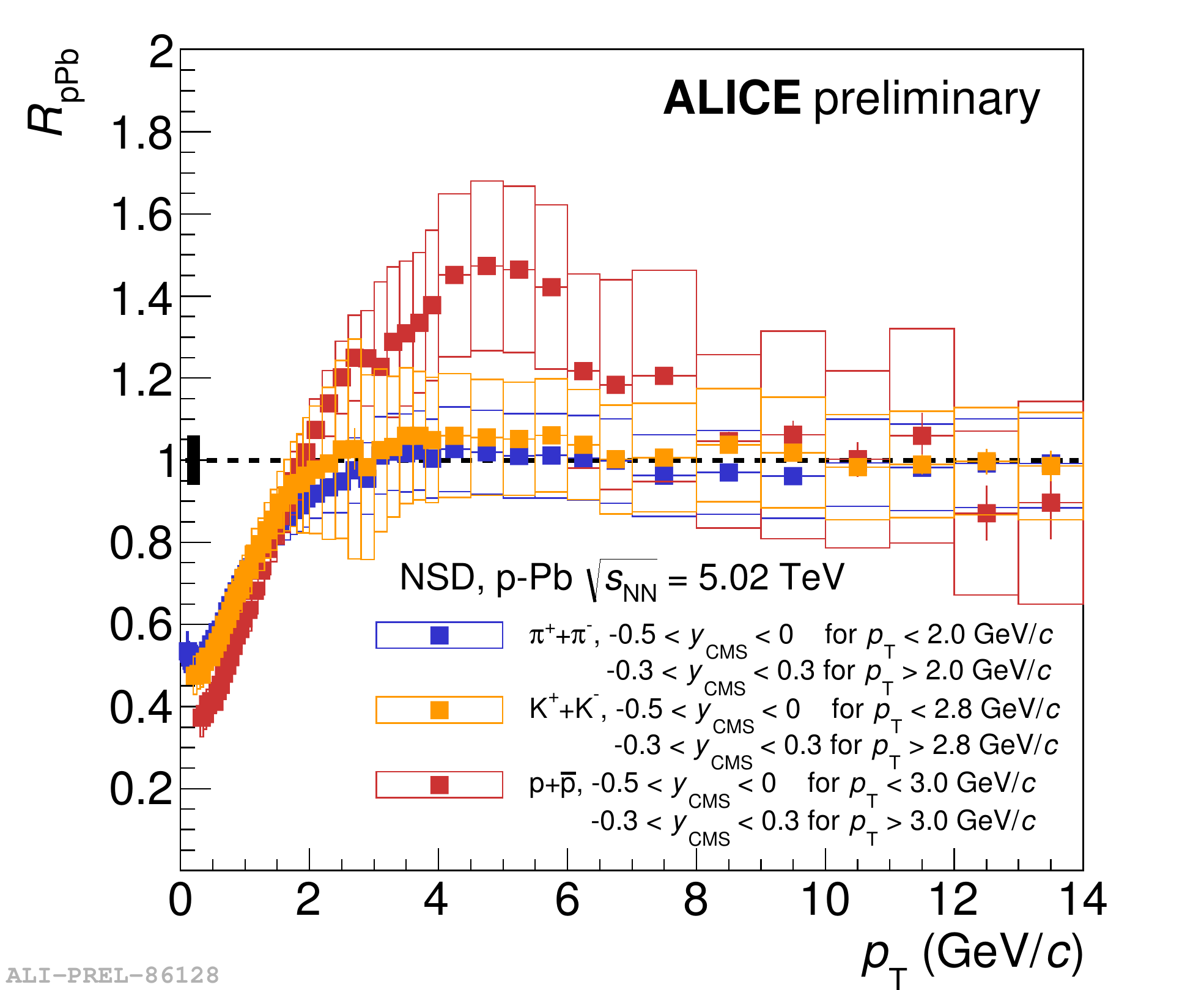}
   \includegraphics[width=0.44\textwidth]{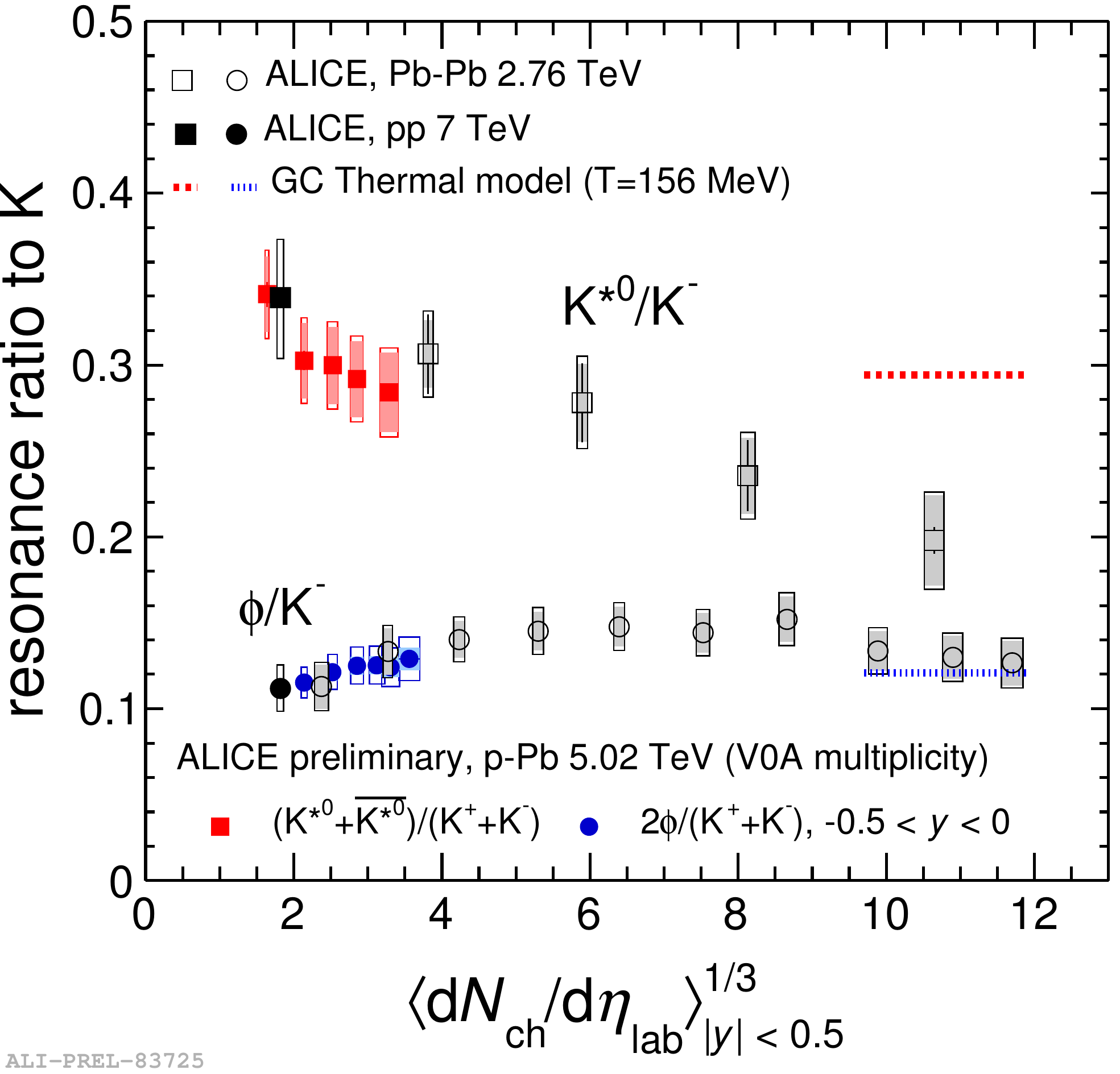}   
   \caption{(Color online).  Nuclear modification factor as a function of \pt for NSD \ppb events. Results for charged pions, kaons and (anti)protons are shown (left). K$^{*0}$ and $\phi$ yields normalized to that for charged kaons as a function of the fireball size, results for different colliding systems are shown: \pbpb, \ppb and  minimum bias \pp collisions (right).}
  \label{fig:pPb:2}
\end{center}
\end{figure}

The nuclear modification factor, \Rppb, is measured in order to study particle species dependence of the nuclear matter effects. \Rppb is defined as:
\begin{equation}
\Rppb = \frac{\text{d}^{2}N_{\text{pPb}}/\text{d}y\text{d}p_{\rm T}}{\left<T_{\rm pPb}\right>\text{d}^{2}\sigma^{\rm INEL}_{\rm pp}/\text{d}y\text{d}p_{\rm T}}
\end{equation}
where, for non-single diffractive (NSD) \ppb collisions the average nuclear overlap function, $\left<T_{\rm pPb}\right>$, is  $0.0983\pm0.0035 \text{\,mb}^{-1}$~\cite{ALICE:2012mj}. In the absence of \pp data at \sppt{5.02}, $\text{d}^{2}\sigma^{\rm INEL}_{\rm pp}/\text{d}y\text{d}p_{\rm T}$ is obtained by interpolating data measured at $\sppt{2.76}$ and at $\sppt{7}$.    The left hand side of Fig.~\ref{fig:pPb:2} shows the identified hadron \Rppb in NSD \ppb events. At high \pt ($>$10\,GeV/$c$), all nuclear modification factors are consistent with unity within systematic and statistical uncertainties.  Around 4\,GeV/$c$, the unidentified charged hadron \Rppb is slightly above unity~\cite{Abelev:2014dsa}, such a enhancement is $\approx$3 times smaller than that observed for the (anti)proton \Rppb. For charged pions and kaons the enhancement is below that of charged particles. This suggests that the barely significant Cronin enhancement observed for inclusive charged particles can be explained  as due to the multiplicity evolution of the (anti)proton spectral shape.

The right hand side of Fig.~\ref{fig:pPb:2} shows the  K$^{*0}$ to charged kaon ratio as a function of the cube root of the average mid-rapidity charged particle density. The \ppb results are compared to analogous measurements in \pp collisions at $\sppt{7}$~\cite{Abelev:2012hy} and \pbpb collisions at $\snnt{2.76}$~\cite{Abelev:2014uua}. In heavy-ion collisions the decreasing trend of K$^{*0}$$/$K with increasing fireball size has been explained as a consequence of a re-scattering of K$^{*0}$ decay daughters in the hadronic phase. It is worth noticing that a similar trend is also observed in \ppb collisions.

\section{Conclusions}
\label{sec:3}

A brief review of recent results on light flavor hadron production in \ppb collisions at $\sqrt{s_{\rm NN}}=$5.02\,TeV has been presented. The results discussed here give an important contribution towards an improved understanding of the origin of the sQGP-like effects discovered in small systems.

Support for this work has been received from CONACYT under the grant No. 260440; and from DGAPA-UNAM under PAPIIT grants IA102515, IN105113, IN107911 and IN108414.

\bibliographystyle{elsarticle-num}

\bibliography{biblio}

\begin{thebibliography}{10}
\expandafter\ifx\csname url\endcsname\relax
  \def\url#1{\texttt{#1}}\fi
\expandafter\ifx\csname urlprefix\endcsname\relax\def\urlprefix{URL }\fi
\expandafter\ifx\csname href\endcsname\relax
  \def\href#1#2{#2} \def\path#1{#1}\fi

\bibitem{Adams:2005dq}
J.~Adams, et~al., Nucl. Phys. {\bf A757} (2005) 102--183.

\bibitem{Adcox:2004mh}
K.~Adcox, et~al., Nucl. Phys. {\bf A757} (2005) 184--283.

\bibitem{Back:2004je}
B.~B. Back, et~al., Nucl. Phys. {\bf A757} (2005) 28--101.

\bibitem{Arsene:2004fa}
I.~Arsene, et~al., Nucl. Phys. {\bf A757} (2005) 1--27.

\bibitem{Schukraft:2011na}
J.~Schukraft, Phil. Trans. Roy. Soc. Lond. {\bf A370} (2012) 917--932.

\bibitem{Heinz:2011kt}
U.~Heinz, C.~Shen, H.~Song, AIP Conf.Proc. {\bf 1441}  766--770.

\bibitem{ABELEV:2013wsa}
B.~B. Abelev, et~al., Phys. Lett. {\bf B726} (2013) 164--177.

\bibitem{Abelev:2013haa}
B.~B. Abelev, et~al., Phys. Lett. {\bf B728} (2014) 25--38.

\bibitem{Schenke:2015aqa}
B.~Schenke, S.~Schlichting, R.~Venugopalan, Phys. Lett. {\bf B747} (2015)
  76--82.

\bibitem{Bozek:2014era}
P.~Bozek, W.~Broniowski, Nucl. Phys. {\bf A926} (2014) 16--23.

\bibitem{Ortiz:2013yxa}
A.~Ortiz, P.~Christiansen, E.~Cuautle, I.~Maldonado, G.~Pai{\'c}, Phys. Rev.
  Lett. 111~({\bf 4}) (2013) 042001.

\bibitem{Armesto:2015ioy}
N.~Armesto, E.~Scomparin, {Heavy-ion collisions at the LHC: a review of the
  results from Run 1}\href {http://arxiv.org/abs/1511.02151}
  {\path{arXiv:1511.02151}}.

\bibitem{Bautista:2015kwa}
I.~Bautista, A.~F. T{\'e}llez, P.~Ghosh, Phys. Rev. {\bf D92}~(7) (2015)
  071504.

\bibitem{Bierlich:2014xba}
C.~Bierlich, G.~Gustafson, L.~L{\"o}nnblad, A.~Tarasov, JHEP {\bf 03} (2015)
  148.

\bibitem{Aamodt:2008zz}
K.~Aamodt, et~al., JINST {\bf 3} (2008) S08002.

\bibitem{Abelev:2014ffa}
B.~B. Abelev, et~al., Int. J. Mod. Phys. {\bf A29} (2014) 1430044.

\bibitem{Adam:2015kca}
J.~Adam, et~al., {Centrality dependence of the $R_{\rm AA}$ of charged pions,
  kaons, and protons in Pb-Pb collisions at $\sqrt{s_{\rm NN}}=2.76$ TeV}\href
  {http://arxiv.org/abs/1506.07287} {\path{arXiv:1506.07287}}.

\bibitem{Abelev:2012hy}
B.~Abelev, et~al., Eur. Phys. J. {\bf C72} (2012) 2183.

\bibitem{Abelev:2014uua}
B.~B. Abelev, et~al., Phys. Rev. {\bf C91} (2015) 024609.

\bibitem{Corke:2010yf}
R.~Corke, T.~Sj{\"o}strand, JHEP {\bf 1103} (2011) 032.

\bibitem{Ortiz:2015cma}
A.~Ortiz, Nucl. Phys. {\bf A943} (2015) 9--17.

\bibitem{Bozek:2013ska}
P.~Bozek, W.~Broniowski, G.~Torrieri, Phys. Rev. Lett. {\bf 111} (2013) 172303.

\bibitem{ALICE:2012mj}
B.~Abelev, et~al., Phys. Rev. Lett. {\bf 110}~(8) (2013) 082302.

\bibitem{Abelev:2014dsa}
B.~B. Abelev, et~al., Eur. Phys. J. {\bf{C74}}~(9) (2014) 3054.

\end{thebibliography}

\end{document}